\documentclass[english,final]{aipproc}
\layoutstyle{6x9}
\usepackage[T1]{fontenc}
\usepackage[latin1]{inputenc}
\usepackage{graphicx}
\usepackage{amssymb}

\providecommand{\tabularnewline}{\\}

 \usepackage{verbatim}

\def\PRC{{\em Phys. Rev.} C}
\def\PRL{{\em Phys. Rev. Lett.}}
\def\NPA{{\em Nucl. Phys.} A}
\def\AP{\em Ann. Phys. (N.Y.)}
\def\PLB{{\em Phys. Lett.} B}
\def\Journal#1#2#3#4{{#1} {\bf #2}, #3 (#4)}

\usepackage{babel}
\begin{document}

\title{New correlations induced by nuclear supersymmetry}

%\maketitle
%\begin{comment}
\author{J. Barea}{address={ICN-UNAM, A.P. 70-543, 04510 México, D.F., México}}

\author{R. Bijker}{address={ICN-UNAM, A.P. 70-543, 04510 México, D.F., México}}

\author{A. Frank}{address={ICN-UNAM, A.P. 70-543, 04510 México, D.F., México}}
%\end{comment}
\begin{abstract}
We show that the nuclear supersymmetry model (n-susy) in its extended
version, predicts correlations in the nuclear structure matrix
elements
which characterize transfer reactions between nuclei that belong
to the same supermultiplet. These correlations are related to the
fermionic generators of the superalgebra and if verified
experimentally can provide a direct test of the model.
\end{abstract}
\maketitle

The extended n-susy model \cite{quartet} correlates spectroscopic
properties of adjacent nuclei. It has been particularly sucessful
in describing the quartet formed by the isotopes $^{194}\textrm{Pt}$,
$^{195}\textrm{Pt}$, $^{195}\textrm{Au}$ and $^{196}\textrm{Au}$
using the $Spin(6)$ limit of the dynamical supersymmetry 
$U_{\nu}(6/12)\otimes U_{\pi}(6/4)$ \cite{quartet,metz}
%corresponding to the group chain\cite{quartet,metz}
%\begin{eqnarray}
%U_{\nu}(6/12)\otimes U_{\pi}(6/4) & \supset & U_{\nu}^{B}(6)\otimes U_{\pi}^{B}(6)\otimes U_{\nu}^{F}(12)\otimes U_{\pi}^{F}(4)\nonumber \\
% & \supset & U_{\nu\pi}^{B}(6)\otimes U_{\nu}^{F}(6)\otimes U_{\nu}^{F}(2)\otimes U_{\pi}^{F}(4)\nonumber \\
% & \supset & U_{\nu\pi}^{BF}(6)\otimes U_{\pi}^{F}(4)\otimes U_{\nu}^{F}(2)\nonumber \\
% & \supset & SO_{\nu\pi}^{BF}(6)\otimes SU_{\pi}^{F}(4)\otimes U_{\nu}^{F}(2)\nonumber \\
% & \supset & Spin(6)\otimes U_{\nu}^{F}(2)\nonumber \\
% & \supset & Spin(5)\otimes U_{\nu}^{F}(2)\nonumber \\
% & \supset & Spin(3)\otimes SU_{\nu}^{F}(2)\nonumber \\
% & \supset & SU(2),\end{eqnarray}
in which the odd-proton is allowed to occupy the $\pi d_{3/2}$
orbit, the odd-neutron occupies the $\nu p_{1/2}$, $\nu p_{3/2}$ and
$\nu f_{5/2}$ orbits, even-even nuclei are described by the
$SO(6)$ limit of the interacting boson model.

In this framework, all states belong to the same irreducible
representation (irrep) of the initial product of supergroups and can
be labeled by different irreps of the subgroups present in the chain
of groups. The states in the quartet are said to form a supermultiplet
and the excitation spectra is obtained using the same hamiltonian.
In this way the excitation energies are correlated. Transitions probabilities
and moments are also correlated through the use of the same
operators.

Both the hamiltonian and the electromagnetic transition
operators are based on the bosonic generators of the
superalgebra, which transform bosons into bosons and fermions
into fermions, inducing transitions inside each
nucleus. In this work we explore the fermionic
generators, which transform bosons into fermions and viceversa.
They are thus associated to transitions among different nuclei,
which we can associate to transfer reactions.

The purpose of this contribution is to show that n-susy establishes
correlations between different transfer reactions. We start with one-nucleon
transfer reactions and then we present the first results, to our knowledge,
of a susy analysis of the two-nucleon transfer reaction $^{198}\textrm{Hg}(\vec{d},\alpha){}^{196}\textrm{Au}$,
in which the wave function correlations of proton-neutron clusters
in the target can be tested.

\section{Correlations in one-nucleon transfer reactions}

The nuclear structure information which can be extracted from
one-nucleon transfer reactions is contained in the so called spectroscopic
intensity, which is the modulus square of the reduced matrix element
of the transfer operator $T^{(J)}$ between the ground state of the
target nucleus in the reaction, and the final state of the residual
nucleus
\begin{equation}
%I=|\langle J;\textrm{F}||T^{(J)}||\textrm{I};0_{gs}\rangle|^{2}.
I=|\langle \alpha_f \, J_f || T^{(J)} || \alpha_i \, J_i \rangle|^{2}.
\end{equation}
It is possible to use a transfer operator deduced from microscopic
asumptions to calculate these quantities, but here we use an
alternative
method based on symmetry considerations. In this case, the transfer
operator can be written as a tensor operator under the subgroups that
appear in the group chain of the dynamical supersymmetry and has the
advantage of giving rise to selection rules and closed expressions
for the spectroscopic intensities \cite{spin6,BI,barea}.

For the one-proton transfer reactions, the transfer tensor operators
read \cite{barea}
\begin{eqnarray}
T_{1,\pi}^{\left\langle \frac{1}{2},\frac{1}{2},-\frac{1}{2}\right\rangle \left(\frac{1}{2},\frac{1}{2}\right)\frac{3}{2}} & = & -\sqrt{\frac{1}{6}}\left(\tilde{s}_{\pi}\times a_{\pi,\frac{3}{2}}^{\dagger}\right)^{\left(\frac{3}{2}\right)}+\sqrt{\frac{5}{6}}\left(\tilde{d}_{\pi}\times a_{\pi,\frac{3}{2}}^{\dagger}\right)^{\left(\frac{3}{2}\right)}\\
T_{2,\pi}^{\left\langle \frac{3}{2},\frac{1}{2},\frac{1}{2}\right\rangle \left(\frac{1}{2},\frac{1}{2}\right)\frac{3}{2}} & = & \sqrt{\frac{5}{6}}\left(\tilde{s}_{\pi}\times a_{\pi,\frac{3}{2}}^{\dagger}\right)^{\left(\frac{3}{2}\right)}+\sqrt{\frac{1}{6}}\left(\tilde{d}_{\pi}\times a_{\pi,\frac{3}{2}}^{\dagger}\right)^{\left(\frac{3}{2}\right)},
\end{eqnarray}
where the upper indices specify the tensorial properties under $Spin(6)$,
$Spin(5)$ and $Spin(3)$. The tensorial character under $Spin(6)$
of $T_{1,\pi}$ implies that it only excites the ground state of the
odd-even nucleus from the ground state of the even-even nucleus. However,
$T_{2,\pi}$ allows the transfer to an excited state in the odd-even
nucleus. The ratio of the intensities for each transfer operator is
given by \cite{barea}
\begin{equation}
R_{1}({\textrm{ee}\rightarrow\textrm{oe}})=\frac{I_{gs\rightarrow exc}}{I_{gs\rightarrow gs}}=0,
\end{equation}
\begin{equation}
R_{2}({\textrm{ee}\rightarrow\textrm{oe}})=\frac{I_{gs\rightarrow exc}}{I_{gs\rightarrow gs}}=\frac{9(N+1)(N+5)}{4(N+6)^{2}},
\end{equation}
where $N$ is taken as the number of bosons in the odd-odd nucleus
of the quartet and ee and oe refer to even-even and odd-even respectively.
In the case of the one-proton transfer $^{194}\textrm{Pt} \rightarrow ^{195}\textrm{Au}$,
the second ratio is $R_{2}=1.12$ ($N=5$), but the relatively small
strength to excited $J=\frac{3}{2}$ states suggests that the operator
$T_{1,\pi}$ can be used to describe the data.

%Due to the $F$-spin symmetry structure of the wave functions of
%the even-odd and odd-odd nuclei in the quartet, they can be linked
%to the wave functions of the even-even and odd-even nuclei, respectively,

Due to the $F$-spin symmetry structure of the wave functions
it is possible to establish the following correlations with the reaction which
involves the even-odd (eo) and odd-odd(oo) nuclei:
\begin{equation}
R_{1}({\textrm{ee}\rightarrow\textrm{oe}})=R_{1}({\textrm{eo}\rightarrow\textrm{oo}})=0,\label{eq:rat1}
\end{equation}
\begin{equation}
R_{2}({\textrm{ee}\rightarrow\textrm{oe}})=R_{2}({\textrm{eo}\rightarrow\textrm{oo}})=\frac{9(N+1)(N+5)}{4(N+6)^{2}}\label{eq:rat2}.
\end{equation}
A different way to understand this result is
through the use of a tensor operator which transforms as a scalar
in the pseudo-$l$ degree of freedom \cite{tesis-Roelof} (upper and
lower indices specify the tensorial properties under $Spin(6)$ and $Spin(5)$,
$Spin(3)$ and $SU(2)$ respectively)
\begin{equation}
P_{\left(0,0\right)0,\frac{1}{2}}^{\left\langle 0,0,0\right\rangle }=\left(\tilde{s}_{\nu}\times a_{\nu,\frac{1}{2}}^{\dagger}\right)^{\left(\frac{1}{2}\right)}-\sqrt{2}\left(\tilde{d}_{\nu}\times a_{\nu,\frac{3}{2}}^{\dagger}\right)^{\left(\frac{1}{2}\right)}+\sqrt{3}\left(\tilde{d}_{\nu}\times a_{\nu,\frac{5}{2}}^{\dagger}\right)^{\left(\frac{1}{2}\right)}.\label{eq:proyector}
\end{equation}
This operator links the wave functions of
the even-odd and odd-odd nuclei in the quartet
to the wave functions of the even-even and odd-even nuclei, respectively. 
The use of this property and the fact that $P$ conmutes with the
one-proton transfer operators $T_{1,\pi}$ and $T_{2,\pi}$, allows
to write the equations (\ref{eq:rat1}) and
(\ref{eq:rat2}) \cite{erice-roelof,future}.
%$R_{1}(\textrm{even-even}\rightarrow\textrm{odd-even})$
%and $R_{2}(\textrm{even-even}\rightarrow\textrm{odd-even})$ to the
%corresponding ratios in the one proton transfer reaction $\textrm{even-odd}\rightarrow\textrm{odd-odd}$\cite{erice-roelof,future}
%\begin{equation}
%R_{1}^{\textrm{\, even-even}\rightarrow\textrm{odd-even}}=R_{1}^{\textrm{\, even-odd}\rightarrow\textrm{odd-odd}}=0,
%\end{equation}
%\begin{equation}
%R_{2}^{\textrm{\, even-even}\rightarrow\textrm{odd-even}}=R_{2}^{\textrm{\, even-odd}\rightarrow\textrm{odd-odd}}=\frac{9(N+1)(N+5)}{4(N+6)^{2}}.
%\end{equation}

We thus find a direct correlation between two different one-proton
reactions, which for the case of $^{195}\textrm{Pt} \rightarrow ^{196} \textrm{Au}$
can be tested experimentally. Very recently this reaction has been
measured \cite{graw} and the experimental data confirms that $T_{1,\pi}$
is capable of describing it, given that the strength to the ground
state of $^{196}\textrm{Au}$ is relatively strong and the excited
state which $T_{2,\pi}$ can excite is not seen populated in the measurements.
This fact confirms also that the correlation seems to apply.

For the one neutron transfer reaction there is a similar correlation.
Let us consider the following transfer tensor operator (labels like
in (\ref{eq:proyector})) \cite{tesis-Roelof}\begin{equation}
T_{(1,0)2,j,\nu}^{(2,0,0)}=\sqrt{\frac{1}{2}}\left(\tilde{s}_{\nu}\times a_{\nu,j}^{\dagger}\right)^{\left(j\right)}-\sqrt{\frac{1}{2}}\left(\tilde{d}_{\nu}\times a_{\nu,\frac{1}{2}}^{\dagger}\right)^{\left(j\right)}\,\,\,\,\, j=\frac{3}{2},\,\frac{5}{2}.\end{equation}
The same argument about the link between the wave functions of
the even-even and the odd-even nuclei permits us to correlate the one
neutron transfer reaction $\textrm{even-even}\rightarrow\textrm{odd-even}$
with the inverse reaction
$\textrm{odd-even}\rightarrow\textrm{even-even}$
\cite{future}.
Taking the following ratios\begin{equation}
R(\textrm{ee} \rightarrow \textrm{oe})=\frac{\left|\left\langle \textrm{oe;}[N_{1},N_{2}]\langle\sigma_{1},\sigma_{2},\sigma_{3}\rangle(1,0)2;J\left\Vert T_{(1,0)2,j,\nu}^{(2,0,0)}\right\Vert \textrm{ee; g.s.}\right\rangle \right|^{2}}{\left|\left\langle \textrm{oe;}[N+1,1]\langle N+1,1,0\rangle(1,0)2;J\left\Vert T_{(1,0)2,j,\nu}^{(2,0,0)}\right\Vert \textrm{ee; g.s.}\right\rangle \right|^{2}},\end{equation}
\begin{equation}
R(\textrm{oe} \rightarrow\textrm{ee})=\frac{\left|\left\langle \textrm{ee};[N_{1},N_{2}]\langle\sigma_{1},\sigma_{2},\sigma_{3}\rangle(1,0)2\left\Vert \tilde{T}_{(1,0)2,j,\nu}^{(2,0,0)}\right\Vert \textrm{oe; g.s.}\right\rangle \right|^{2}}{\left|\left\langle \textrm{ee};[N+1,1]\langle N+1,1,0\rangle(1,0)2\left\Vert \tilde{T}_{(1,0)2,j,\nu}^{(2,0,0)}\right\Vert \textrm{oe; g.s.}\right\rangle \right|^{2}},\end{equation}
we find the following relations
\[
R(\textrm{oe} \rightarrow \textrm{ee})=R(\textrm{ee} \rightarrow \textrm{oe}) \textrm{ for } N_{2}=1,
\]
\[
R(\textrm{oe} \rightarrow \textrm{ee})=R(\textrm{ee} \rightarrow \textrm{oe}) \frac{(N+1)(N_{\nu}+1)}{(N_{\pi}+1)}\textrm{ for }N_{2}=0,
\]

In table \ref{tab:one-neutron-transfer} we show these ratios and
quote the experimental values \cite{exp} and the calculated ones for the isotopes
$^{194}\textrm{Pt}$ and $^{195}\textrm{Pt}$. We can observe a consistence
between the experimental and calculated values, but clearly more experimental
work is needed to confirm if these correlations hold.\renewcommand{\arraystretch}{1.5}{%
\begin{table}

\caption{Intensity ratios for one-neutron transfer reactions.
\label{tab:one-neutron-transfer}}

\begin{tabular}{cccc}
\hline 
$[N_{1},N_{2}]\langle\sigma_{1},\sigma_{2},\sigma_{3}\rangle$&
$R_{ee\rightarrow oe}$&
\begin{tabular}{cc}
\multicolumn{2}{c}{$^{194}\textrm{Pt}\rightarrow{}^{195}\textrm{Pt}$}\tabularnewline
\begin{tabular}{c} calc. \tabularnewline \tabularnewline \end{tabular}
&
\begin{tabular}{cc}
\multicolumn{2}{c}{exp.}\tabularnewline
$j = \frac{3}{2}$ & $j = \frac{5}{2}$\tabularnewline
\end{tabular}
\tabularnewline
\end{tabular}&
\begin{tabular}{cc}
\multicolumn{2}{c}{$^{195}\textrm{Pt}\rightarrow{}^{194}\textrm{Pt}$}\tabularnewline
\begin{tabular}{c} calc. \tabularnewline \tabularnewline \end{tabular}
&
\begin{tabular}{c} exp. \tabularnewline \tabularnewline \end{tabular}
\tabularnewline
\end{tabular}\tabularnewline
\hline
$[N+2]\langle N+2,0,0\rangle$&
$\frac{2(N+4)}{(N+1)(N+3)(N+6)}$&
\begin{tabular}{cc}
0.034&
\begin{tabular}{cc} 0.264 & 0.052 \tabularnewline \end{tabular}
\tabularnewline
\end{tabular}&
\begin{tabular}{cc}
0.511&
-------\tabularnewline
\end{tabular}\tabularnewline
$[N+2]\langle N,0,0\rangle$&
$\frac{N(N+4)(N+5)}{2(N+2)(N+3)(N+6)^{2}}$&
\begin{tabular}{cc}
0.033&
\begin{tabular}{cc} ------- &  ------- \tabularnewline \end{tabular}
\tabularnewline
\end{tabular}&
\begin{tabular}{cc}
0.498&
-------\tabularnewline
\end{tabular}\tabularnewline
$[N+1,1]\langle N,0,0\rangle$&
$\frac{N²(N+5)}{2(N+2)(N+6)²}$&
\begin{tabular}{cc}
0.148&
\begin{tabular}{cc} 0.087 &  ------- \tabularnewline \end{tabular}
\tabularnewline
\end{tabular}&
\begin{tabular}{cc}
0.148&
-------\tabularnewline
\end{tabular}\tabularnewline
\hline
\end{tabular}
\end{table}
}

\section{Correlations in two-nucleon transfer reactions}

In contrast to the one-nucleon transfer reactions in which the single
particle structure of nuclear states is examinated, two-nucleon transfer
reactions are very sensitive to the correlation between the transfered
nucleons. As a consequence, this kind of transfer reactions supply
a stringent test for the nuclear wave functions.

Two factors determine the strength of a two nucleon transfer reaction,
which are related to the two-nucleon fractional parentage coefficients.
On the one hand it depends on how similar the state of $A+2$ nucleons is
to the state of $A$ plus two additional nucleons, and on
the other, if the correlation of these two nucleons in the $A+2$
state is similar to the correlation in the state of the light nucleus.
The nuclear structure information that can be extracted from a model
is related to both factors. We shall follow the formalism developed
by Glendenning \cite{Glendenning} and briefly sketch the main ingredients
as applied to the reaction $^{198}\textrm{Hg}(\vec{d},\alpha){}^{196}\textrm{Au}$,
which has been measured recently to study the odd-odd nucleus $^{196}\textrm{Au}$ \cite{graw}.

\begin{comment}
The differential cross section is given by\begin{equation}
\frac{d\sigma}{d\Omega}=\frac{m_{\alpha}m_{\beta}}{(2\pi\hbar^{2})^{2}}\frac{k_{\beta}}{k_{\alpha}}\frac{\hat{J}_{B}}{\hat{J}_{A}}...,\end{equation}
where $m_{\alpha}$is ... 
\end{comment}
The nuclear structure information is contained in the spectroscopic
strengths $G_{LJ}$. This quantities are written as\begin{equation}
G_{LJ}=\left|\sum_{Nj_{\nu}j_{\pi}}\beta_{j_{\nu}j_{\pi}}G_{NLJ}^{j_{\nu}j_{\pi}}\right|^{2},\end{equation}
where\begin{eqnarray}
\beta_{j_{\nu}j_{\pi}} & = & \left\langle ^{196}\textrm{Au;}J\left\Vert T_{j_{\nu}j_{\pi}}^{(J)}\right\Vert {}^{198}\textrm{Hg};0_{gs}\right\rangle \\
G_{NLJ}^{j_{\nu}j_{\pi}} & = & \sqrt{3}\hat{L}\hat{j}_{\nu}\hat{j}_{\pi}\left\{ \begin{array}{ccc}
l_{\nu} & \frac{1}{2} & j_{\nu}\\
l_{\pi} & \frac{1}{2} & j_{\pi}\\
L & 1 & J\end{array}\right\} \left\langle n\,0\, N\, L\left|n_{\nu}\, l_{\nu}\, n_{\pi}\, l_{\pi}\right.;L\right\rangle .\end{eqnarray}
For the two nucleon transfer operator $T_{j_{\nu}j_{\pi}}^{(J)}$
we have chosen the simplest possible form:\begin{equation}
T_{j_{\nu}j_{\pi}}^{(J)}=\alpha_{j_{\nu}}\left(a_{j_{\nu}}^{\dagger}\times a_{j_{\pi}}^{\dagger}\right)^{(J)}.\end{equation}
The parameters $\alpha_{j_{\nu}}$ are determined by using a least
square fit of the spectroscopic strengths to the experimental values.
%These experimental values are extracted from the measured cross sections
%$\sigma_{+}$ and $\sigma_{-}$ (for spin up and spin down) in comparing
%the differential cross sections and asymmetries\begin{eqnarray}
%\frac{d\sigma}{d\Omega} & = & \frac{\sigma_{+}+\sigma_{-}}{2}\\
%A_{y} & = & \frac{\sigma_{+}-\sigma_{-}}{3P_{y}}\left(\frac{d\sigma}{d\Omega}\right)^{-1},\end{eqnarray}
%with the values obtained with the code CHUCK3\cite{CHUCK3}, which
%calculates theoretical $\frac{d\sigma}{d\Omega}^{*}$ and $A_{y}^{*}$
%from theoretical differential cross sections $\sigma^{LJ}$, vector
%analysing powers $A_{y}^{LJ}$ and tensor analysing powers $A_{yy}^{LJ}$
%values for the respective $LJ$ transfers:\begin{eqnarray}
%\frac{d\sigma}{d\Omega}^{*} & = & \sum_{LJ}G_{LJ}\sigma^{LJ}\left[1+\frac{P_{yy}A_{yy}^{LJ}}{2}\right]\\
%A_{y}^{*} & = & \sum_{LJ}G_{LJ}\sigma^{LJ}A_{y}^{LJ}\left[1+\frac{P_{yy}A_{yy}^{LJ}}{2}\right]\left(\frac{d\sigma}{d\Omega}^{*}\right)^{-1}.\end{eqnarray}
To compare the experimental and calculated spectroscopic strengths
we have choosen seven states of reference, each one
corresponding to
each of the seven possible $LJ$ transfers that the n-susy model
allows,
and we have calculated for each $LJ$ the transfer ratio\begin{equation}
R_{LJ}=\frac{G_{LJ}}{G_{LJ}^{ref}},\end{equation}
where $G_{LJ}^{ref}$ is the spectroscopic strength for the reference
state for a particular $LJ$ tranfer. %
\begin{figure}

\caption{Ratios of spectroscopic strengths. The first column in each frame
correspond to states with $\left(\frac{3}{2},\frac{1}{2}\right)$
$Spin(5)$ labels and the second with $\left(\frac{1}{2},\frac{1}{2}\right)$
labels, respectively. Each row from the bottom to the top corresponds
to states with labels $[6,0]\langle6,0\rangle(13/2,1/2,1/2)$, $[5,1]\langle5,1\rangle(11/2,1/2,1/2)$
and $[5,1]\langle5,1\rangle(11/2,3/2,1/2)$ for the groups $U_{\nu\pi}^{BF}(6)$,
$SO_{\nu\pi}^{BF}(6)$ and $Spin(6)$.\label{fig:Ratios-of-spectroscopic}}

\includegraphics[%
  width=1.0\textwidth,
%  keepaspectratio]{/home/barea/cocoyoc04/G-all-gris.eps}
  keepaspectratio]{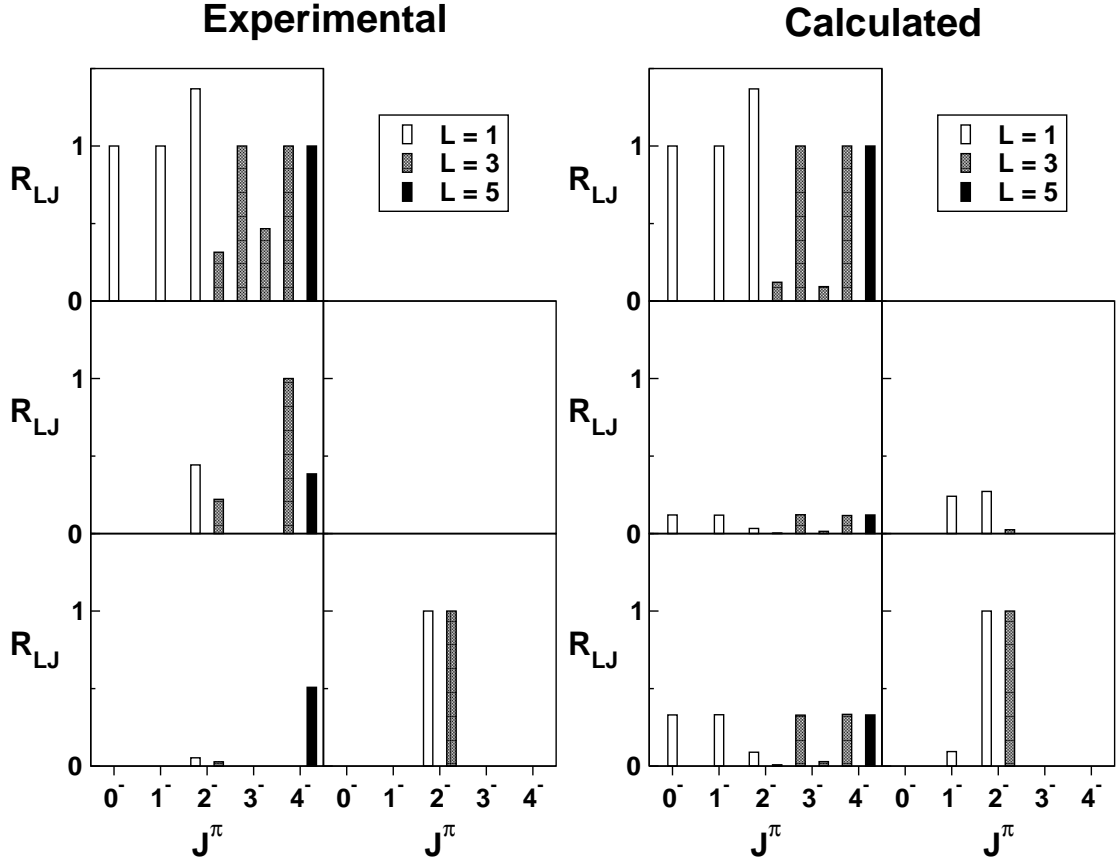}
\end{figure}
In figure \ref{fig:Ratios-of-spectroscopic} we show the experimental
and calculated ratios $R_{LJ}$. We observe from this figure that
the comparison is quite reasonable if we take into account the simple form we adopted
for the two nucleon transfer operator. In concluding this section
we can say that the n-susy model can reproduce remarkably well the main
features of the neutron-proton correlations which are present in the
nuclear states involved.

\section{Summary}

We have found new correlations predicted by extended n-susy. These
correlations relate spectroscopic strengths of different one-nucleon
transfer reactions between the nuclei which are described by the model.
We focused on the supermultiplet formed by $^{194}\textrm{Pt}$, $^{195}\textrm{Pt}$,
$^{195}\textrm{Au}$ and $^{196}\textrm{Au}$ whose spectroscopic
properties have been previously described in terms of this model. 
We have shown that these correlations are partially fulfilled
with the availaible experimental data, but clearly more experimental
work is necessary to test their full validity.

We have calculated the spectroscopic strengths associated to the two
nucleon transfer reaction $^{198}\textrm{Hg}(\vec{d},\alpha){}^{196}\textrm{Au}$,
which was recently measured. The comparison between the calculated
and experimental data shows good agreement considering the
simple form adopted for the two nucleon transfer operator.

We plan to search for other experimental examples to which extended
n-susy and its correlations can be applied, eventually relaxing the constraints
set by dynamical symmetry \cite{erice-frank,frank-isacker-warner}.
We wish to emfasize that nuclear susy may be a model whose range of
applicability is wider than was previously realized and which may
lay the foundations of a new and unifying point of view in nuclear
structure.

\section{Acknowledgments}

This work was supported by CONACyT. We are grateful to G. Graw for
sharing the new experimental data on the transfer reactions $^{195}\textrm{Pt}({}^{3}\textrm{He,d}){}^{196}\textrm{Au}$
and $^{198}\textrm{Hg}(\vec{d},\alpha){}^{196}\textrm{Au}$ prior
to publication. Many enlighting discussions with J. G\'omez-Camacho and P.
Van Isacker are acknowledged.

\end{document}